\documentclass[letterpaper, 9pt, conference]{ieeeconf}
\IEEEoverridecommandlockouts
\usepackage{blindtext, graphicx}
\usepackage{tikz}
\usepackage[siunitx, europeanresistors,americaninductors]{circuitikz}
\usepackage{bm}
\usepackage{array}
\usepackage{amsmath}
\allowdisplaybreaks[4]
\usepackage{xcolor}
\usepackage{lipsum}
\usepackage{float}
\usepackage{cite}
\usepackage{url}
\usepackage{hyperref}

\usepackage{amssymb}
\DeclareMathAlphabet{\pazocal}{OMS}{zplm}{m}{n}
\newcommand{\Nb}{\mathcal{N}}
\newcommand{\Tb}{\mathrm{T}}
\newcommand{\Rb}{\mathrm{R}}

\newcommand{\Eb}{\mathcal{E}}
\newcommand{\Sb}{\ensuremath{\mathrm{S}}}

\newcommand{\diag}{\ensuremath{\mathrm{diag}}}

\title{\Huge Assessing the Optimality of LinDist3Flow\\ for Optimal Tap Selection of Step Voltage Regulators\\ in Unbalanced Distribution Networks}
\author{}
\author{Krishna Sandeep Ayyagari$^{\star}$, Sherin Ann Abraham$^{\dagger}$, Yiyun Yao$^{\dagger}$, Shibani Ghosh$^{\dagger}$, \\  Francisco Flores-Espino$^{\dagger}$, Adarsh Nagarajan$^{\dagger}$,  Nikolaos Gatsis$^{\star}$ 
	\thanks{$^{\star}$Department of Electrical Engineering, The University of Texas at San Antonio, TX 78249. $^{\dagger}$National Renewable Energy Laboratory
Golden, Colorado, $^{\star}$Emails:\{krishnasandeep.ayyagari, nikolaos.gatsis\}@utsa.edu, $^{\dagger}$Emails:\{sherinann.abraham, yiyun.yao, shibani.ghosh, francisco.flores, adarsh.nagarajan\}@nrel.gov.
This material is based upon work supported by the National Science Foundation under Grants  1847125, 2115427, and by the U.S. Department of Energy Office of Energy Efficiency and Renewable Energy Solar Energy Technologies Office. This work was authored in part by the National Renewable Energy Laboratory, operated by Alliance for Sustainable Energy, LLC, for the U.S. Department of Energy (DOE) under Contract No. DE-AC36-08GO28308. The views expressed in the article do not necessarily represent the views of the DOE or the U.S. Government. The U.S. Government retains and the publisher, by accepting the article for publication, acknowledges that the U.S. Government retains a nonexclusive, paid-up, irrevocable, worldwide license to publish or reproduce the published form of this work, or allow others to do so, for U.S. Government purposes.}%
}

\begin{document}
	
	\maketitle
	\thispagestyle{empty}
	\pagestyle{empty}

	\begin{abstract}
     The adoption of distributed energy resources such as photovoltaics (PVs) has increased dramatically during the previous decade. 
     The increased penetration of PVs into distribution networks (DNs) can cause voltage fluctuations that have to be mitigated. One of the key utility assets employed to this end are step-voltage regulators (SVRs). It is desirable to include tap selection of SVRs in optimal power flow (OPF) routines, a task that turns out to be challenging because the resultant OPF problem is nonconvex with added complexities stemming from accurate SVR modeling. While several convex relaxations based on semi-definite programming (SDP) have been presented in the literature for optimal tap selection, SDP based schemes do not scale well and are challenging to implement in large-scale planning or operational frameworks. 
     This paper deals with the optimal tap selection (OPTS) problem  for  wye-connected SVRs using linear approximations of power flow equations. Specifically, the $\textit{LinDist3Flow}$ model is adopted and the effective SVR ratio is assumed to be continuous--enabling the formulation of a problem called $\textit{LinDist3Flow-OPTS}$, which amounts to a linear program. The scalability and optimality gap of $\textit{LinDist3Flow-OPTS}$ are evaluated with respect to existing SDP-based and nonlinear programming techniques for optimal tap selection in three standard feeders, namely, the IEEE 13-bus, 123-bus, and 8500-node DNs. For all DNs considered, $\textit{LinDist3Flow-OPTS}$ achieves an optimality gap of approximately $1\%$ or less while significantly lowering the computational burden. 
	\end{abstract}
	
	\begin{keywords}
	Power distribution networks, step-voltage regulators, optimal power flow, linear approximations
	\end{keywords}


\section{Introduction}
To combat climate change while reaping health and economic benefits, a 100 $\%$ renewable energy power system adoption by 2045, if not sooner, is required~\cite{LA100_techreport}. Furthermore, as the cost of domestic-scale distributed energy resources (DERs) such as photovoltaic (PV) systems and electric vehicles has decreased, DER adoption is expected to accelerate in the near future. 

The integration of DERs into distribution networks results in time-varying active power injections and frequently reversing power flow, causing frequent voltage fluctuations. As a result, maintaining bus voltage magnitudes within desirable levels is critically challenged. In response, the settings of utility owned equipment such as step-voltage regulators (SVRs) need to be re-adjusted to cope with the reversing power flow. Wear and tear of SVRs due to excessive tap changes as a result of DER fluctuation is also a concern~\cite{Alsaleh_2021, Shibani_2019,nagarajan2018studies}. 

To avoid the aforementioned issues, utilities employ tap selection into their optimal power flow (OPF) routines. However, incorporating SVR taps as decision variables in OPF is challenging. Specifically, the nonconvex equality constraints of the multi-phase power flow equations, combined with the discrete mechanical settings of SVRs, in general render the OPF into a mixed-integer nonlinear programming (MINLP) problem. The literature pertaining to the optimal tap selection problem is reviewed next. 

The work in~\cite{Wu2017} develops a mixed-integer second-order cone programming (MISOCP) model for tap selection of on-load tap-changer transformers, where the trilinear scalar constraint in transformer taps and voltages is converted to an exact mixed-binary linear constraint via binary expansion and big-M approaches.  
The computational performance of the MISOCP model is further evaluated considering single- and multi-period optimization problems in~\cite{Savasci_2021} . 
The works in~\cite{Wu2017} and \cite{Savasci_2021} focus on the single-phase DNs to circumvent the modeling complexity of multi-phase DNs. However, because real-world DNs are intrinsically unbalanced and require a complete multi-phase modeling to deliver a reasonable result, these approaches may not ensure optimal performance. 

To tackle the complexity of unbalanced multi-phase DN operation, approximations or convex relaxations of the power flow equations have been proposed in the literature. 
Specifically, the work in~\cite{gan2014convex} introduces an approximate linear power flow $\textit{LPF}$ model by ignoring loss components and further assuming that the voltages across phases have similar magnitude and differ by an angle~$120^\circ$, i.e., they are balanced. Leveraging~\cite{gan2014convex}, the work in~\cite{Jha2019} develops a bi-level framework for coordinating SVRs and PV inverters, where 
discrete taps for SVRs and PV setpoints are decided via mixed-integer linear programming (MILP) in the first level, and the resultant PV setpoints are revised using  nonlinear programming (NLP) in the second level by fixing the decisions of SVRs to the ones computed in the first level. The optimality of the bi-level method in~\cite{Jha2019} is not evaluated against convex relaxations in the context of optimal tap selection, while the formulations are confined to small-scale DNs. 

Focusing on OPF problems for PV reactive power dispatch, the works in~\cite{RobbinsOPF} and \cite{Arnold_2016} have shown that approximating loss terms as constants that are periodically updated based on the desired operating point significantly improves the optimality of the $\textit{LPF}$ model. By generalizing the assumptions on balanced voltages and higher-order loss terms,~\cite{RobbinsOPF} and \cite{Arnold_2016} extend the $\textit{LPF}$ model of~\cite{gan2014convex} to the $\textit{LinDist3Flow}$ model. Moreover, it has recently been demonstrated in ~\cite{Inaolaji_2021} that the $\textit{LinDist3Flow}$ model exhibits smaller error when compared to other linear models, e.g.,~\cite{Dhople_2015} when used for solving the power flows in unbalanced DNs.

The works in~\cite{Li_2018, Li_2020} introduce a linearization technique to represent SVRs in admittance-based OPF. Specifically, to approximate nonlinearities arising from modeling SVRs taps in the bus admittance matrix, Taylor series expansion is performed around a known tap position, while linearizations of the rectangular form of power flow equations are adopted as well. 

The work in~\cite{Garcia_2018} adopts the nonlinear power flow equations and focuses on open-delta SVRs. The $\textit{OPTS}$ problem is solved by an iterative process based on active-set sequential quadratic programming algorithm, which falls into the category of NLP. It is worth noting that NLP formulations are computationally intensive and cannot in general provide certificates of global optimality. 

Another line of works explores convex relaxations
based on  semidefinite programming (SDP)~\cite{Robbins2016, Liu_2018, Liu_2019, Hafez_PSCC, Hafez_OPF_SVR, Alsaleh_2021}.  A full SDP framework for admittance-based OPF is developed in~\cite{Robbins2016} for optimal tap selection considering wye-connected SVRs. The trilinear matrix constraint in taps and voltages is relaxed to a linear constraint by confining the diagonal of the secondary-side voltage to the minimum and maximum tap ratio range. By exploiting the chordal SDP relaxation of the admittance-based OPF, the trilinear matrix constraint in taps and voltages is relaxed to a linear semidefinite matrix constraint in~\cite{Liu_2018} under the assumption that taps on each phase of the SVR are equal. The work in~\cite{Hafez_PSCC} develops chordal SDP relaxation of the admittance-based OPF considering wye, closed-delta, and open-delta SVRs with independent tap operation. In~\cite{Hafez_OPF_SVR}, a more accurate and computationally stable formulation is proposed which allows for tap selection of wye, closed-delta, and open-delta SVRs using the branch flow-based OPF, McCormick envelopes, and a phase separation assumption on the secondary voltage of the SVR. The works in~\cite{Robbins2016, Liu_2018, Hafez_PSCC, Hafez_OPF_SVR} assume continuous values for the effective regulation ratio of SVRs. A branch flow-based SDP model to optimally dispatch SVRs and PV inverters considering discrete decisions of SVRs is developed in~\cite{Alsaleh_2021}, which renders the overall formulation into a mixed-integer SDP (MISDP) problem and is solved using Generalized Benders Decomposition.

SDP approaches are challenging to implement in practice for large DNs. The challenge is compounded in the presence of multiple time periods, since the number of variables in SDP grows significantly with the number of buses~\cite{RobbinsOPF}. Additionally, including devices with discrete decisions, e.g., SVR taps, battery energy storage devices, and capacitor banks, into SDP approaches or extending these approaches for planning problems pertaining to optimal SVR placement in unbalanced DNs is challenging. The reason is that it is hard to solve the resulting MISDP problems, given the limited performance of off-the-shelf SDP solvers for large-scale formulations. The previously mentioned problems are critical for instance in planning studies in the context of $100\%$ renewable energy integration~[\citenum{LA100_techreport}, Ch. VII]. Overall, there is a need to develop computationally attractive schemes for tap selection in unbalanced DNs. 

Although several linearization techniques techniques have been proposed in the literature---see e.g.,~\cite{Garces_2016, Bernstein_2017, Huang_2021, Girigoudar_2021}---the $\textit{LinDist3Flow}$ has been extensively used to solve for PV reactive power setpoints and other DN scheduling tasks demonstrating encouraging results. However, the $\textit{LinDist3Flow}$ model has not been fully tested in the context of optimal tap selection. Therefore, evaluating the performance of $\textit{LinDist3Flow}$ model for optimal tap selection problems remains open and the tradeoff between optimality and computational effort needs to be investigated.  

In this context, this paper investigates the optimal tap selection with $\textit{LinDist3Flow}$ power flows enabling the solution of the problem with less computational burden and a reasonable optimality gap when benchmarked against SDP relaxations. The focus is on wye-connected SVRs. Following a similar procedure in~\cite{Robbins2016}, the trilinear matrix constraint in taps and voltages is relaxed to a linear constraint by confining the primary-side voltage to the minimum and maximum tap ratio range. The resulting $\textit{LinDist3Flow}$ optimal tap selection problem is called  $\textit{LinDist3Flow-OPTS}$ and amounts to a linear programming (LP) formulation. The $\textit{LinDist3Flow-OPTS}$ problem is extensively tested on the IEEE 13-bus, 123-bus, and 8500-node distribution feeders, which include one-, two-, and three-phase wye-connected SVRs respectively. Additionally, detailed numerical comparisons in terms of solution quality, computation time, and linearization accuracy are provided with respect to SDP approaches of~\cite{Robbins2016, Liu_2018, Hafez_OPF_SVR} as well as traditional NLP formulations. In particular, $\textit{LinDist3Flow-OPTS}$
provides solutions with  \emph{approximately} $1\%$ \emph{optimality gap}  and at a much lower computational
cost compared to the conventional NLP and SDP approaches for the IEEE 8500-node DN.

The remainder of the paper are organized as follows. Section II describes the nonconvex branch-flow model for optimal tap section $\textit{BF-OPTS}$. The formulation of $\textit{LinDist3Flow-OPTS}$ is detailed in Section III. Section IV provides the case studies on the standard IEEE feeders and compares the performance of $\textit{LinDist3Flow-OPTS}$ with NLP and SDP-based approaches. The paper concludes in Section V.

$\textit{Notation}$:  The
notation $\bar{(.)}$ is used to denote the complex conjugate transpose of $(.)$. Operator $\mathrm{Re}(.)$ returns the real part of complex number. The notation $(.)^{*}$ denotes the complex conjugate of $(.)$. For a vector $(.)$, $\diag$  returns the square matrix with elements of $(.)$ on the main diagonal; for a square matrix $(.)$, $\diag$ returns the vector with elements from the main diagonal of $(.)$.

\section{Distribution Network Model and\\ Branch Flow Optimal Tap Selection}
Consider a multi-phase distribution network (DN) which is modeled by a directed tree graph ($\Nb$, $\Eb$) with buses collected in set $\Nb$, where $\Nb := \{1,\ldots, N\}\cup \{\Sb\}$. Set $\Eb \subseteq \Nb \times \Nb$ collects the edges, which are all pointing away from the root. The root is denoted by node $\Sb$, which models the secondary of the substation transformer and is considered the slack bus with known constant voltage $v_{\Sb}\in\mathbb{C}^3$. The set of edges represents the network series elements including distribution lines and transformers collected in set $\Eb_{\Tb}$, and step-voltage regulators (SVRs) included in $\Eb_{\Rb} $; that is, $\Eb = \Eb_{\Tb} \bigcup \Eb_{\Rb} $. The set of available phases at bus $ n \in \Nb_{+} := \Nb \setminus \{\Sb\}$, may include one, two, or three phases. For the sake of exposition, we assume that the full set of three phases $\{a, b, c\}$ are present in all buses $n \in \Nb$ and distribution lines. The extension to networks with missing phases can be carried out using more elaborate notation, and the numerical results include tests on networks with missing phases. 
Notations $(n, m)$ and $n \rightarrow m$ are used interchangeably for edges.

\subsection{Modeling of Series Elements}
Let $v_{n}$, $i_{nm} \in \mathbb{C}^{3}$, and $Z_{nm} \in \mathbb{C}^{3 \times 3}$ respectively denote the vector of phase voltages at bus $n \in \Nb$, the vector of line currents, and the series impedance of distribution line $(n,m)\in \Eb_{\Tb}$ (neglecting the shunt admittance)~\cite{Kersting_2018} or the inverse of the per-unit shunt admittance for a grounded-wye grounded-wye transformer $(n,m)\in \Eb_{\Tb}$~\cite{Zbushafez}. 
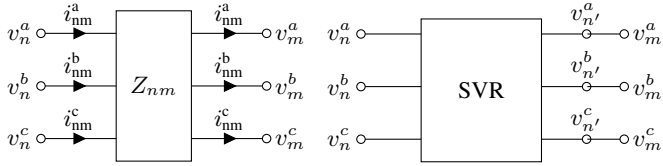
\begin{figure}[t]  
\begin{tikzpicture}[american voltages, baseline=(current bounding box.center)]
  \ctikzset{ label/align = straight }
    \draw  (-5,-0.2) node[left]{$v^{c}_{n}$} to[short,i=${i}^{\text{c}}_{\text{nm}}$,o-] (-4,-0.2);
     \draw (-5,0.5) node[left] {$v^{b}_{n}$} to[short,i=${i}^{\text{b}}_{\text{nm}}$,o-] (-4,0.5);
     \draw (-5,1.2)node[left]{$v^{a}_{n}$} to[short,i=${i}^{\text{a}}_{\text{nm}}$,o-] (-4,1.2);
     \draw (-4,-0.5) rectangle (-3,1.5) node[pos=.5] {${Z}_{nm}$};
     \draw (-3,1.2) to [short,i=${i}^{\text{a}}_{\text{nm}}$, -o] (-2,1.2) node[right] {$v^{a}_{m}$};
     \draw (-3,0.5) to [short,i=${i}^{\text{b}}_{\text{nm}}$, -o] (-2,0.5) node[right] {$v^{b}_{m}$};
     \draw (-3,-0.2) to [short,i=${i}^{\text{c}}_{\text{nm}}$, -o] (-2,-0.2) node[right] {$v^{c}_{m}$};
    \end{tikzpicture}
    \begin{tikzpicture}[american voltages, baseline=(current bounding box.center)]
  \ctikzset{ label/align = straight }
   \draw (-1,1.2) node[left]{$v^{a}_{n}$} to[short,o-] (-0.2,1.2);
     \draw (-1,0.5) node[left]{$v^{b}_{n}$} to[short,o-] (-0.2,0.5);
      \draw (-1,-0.2)node[left]{$v^{c}_{n}$} to[short,o-] (-0.2,-0.2);
      \draw (-0.2,-0.5) rectangle (1.4,1.4) node[pos=.5] {$\textrm{SVR}$};
      \draw (1.4,1.2) to [short, -o] (2.0,1.2) node[above] {$v^{a}_{n'}$};
       \draw (1.4,0.5) to [short, -o] (2.0,0.5) node[above] {$v^{b}_{n'}$};
        \draw (1.4,-0.2) to [short, -o] (2.0,-0.2) node[above] {$v^{c}_{n'}$};
         \draw (2.01,-0.2) to [short, -o] (2.5,-0.2) node[right] {$v^{c}_{m}$};
         \draw (2.01,0.5) to [short, -o] (2.5,0.5) node[right] {$v^{b}_{m}$};
         \draw (2.01,1.2) to [short, -o] (2.5,1.2) node[right] {$v^{a}_{m}$};
    \end{tikzpicture}
    \caption{Modeling of series element $(n,m)\in \Eb_{\Tb}$ (left) and step-voltage regulator $(n,n')\in \Eb_{\Rb}$ in series with distribution line (right)~\cite{Zbushafez}}
    \label{fig:serieslement}
    \end{figure}
The voltage drop across edge $(n,m) \in \Eb_{\Tb}$ is described by Ohms' law 
\begin{align}
    v_{n} &= v_{m} +Z_{nm} i_{nm}, (n,m) \in \Eb_{\Tb} 
    \label{eq:ohmslaw}
\end{align}

A step-voltage regulator is a series element  which is installed either at the feeder head/substation or along the feeder to regulate voltages of downstream buses. The present work deals with three-, two-, or one-phase grounded-wye grounded-wye SVRs.  The SVR is modeled as shown in~Fig.\ref{fig:serieslement} (right)~\cite{Zbushafez}, i.e., the SVR is in series with a distribution line. Specifically, the SVR device is between buses $n$ and $n'$, that is, $(n,n') \in \Eb_{\Rb}$,
and the edge from $n^{\prime}$ to $m$ corresponds to the distribution line $(n^{\prime}, m) \in \Eb_{\Tb}$ whose voltage drop is described by~\eqref{eq:ohmslaw}. The aforementioned arrangement is typical in DNs ($n'\in\Nb_{+}$ is not a fictitious node). Other
than edges $(n \ n^{\prime})$ and $(n^{\prime}\ m)$ no other edges or current sources are connected to bus $n^{\prime}$. The SVR primary is thus connected to bus $n$ and the secondary is connected to  bus $n^{\prime}$. Furthermore, the SVR is assumed to be ideal, i.e., the series
impedance of the constituent autotransformers is negligible~\cite{Hafez_OPF_SVR}. 

The voltage and current relationships for a type-B SVR on edge $(n,n^{\prime}) \in \Eb_{\Rb}$ are given by
\begin{align}
  {v}_{n} &= {A}_{nn^{\prime}}{v}_{n^{\prime}}  \label{eq:voltage gain} \\
  i_{n n^{\prime}} & = \bar{A}^{-1}_{nn'}i_{n^{\prime}m} \label{eq:current gain}\\
  {A}_{nn^{\prime}} & = \begin{bmatrix}
  r^{a}_{nn^{\prime}} & 0 & 0 \\
  0 & r^{b}_{nn^{\prime}} & 0 \\
  0 & 0 & r^{c}_{nn^{\prime}}
                \end{bmatrix} \label{eq:gain matrix}
\end{align}
where ${A}_{nn^{\prime}}$ is the voltage gain matrix of the SVR. The tap position of the SVR determines the effective regulator ratio $r^{\phi}_{nn^{\prime}}$. Since each tap changes results in $\frac{5}{8}\%$ or $0.00625~\text{p.u}$ change in voltage, the effective regulator ratio is given by $r^{\phi}_{nn^{\prime}} = 1 - 0.00625 t^{\phi}_{nn^{\prime}}$ where the tap $ t^{\phi}_{nn^{\prime}}$ varies in $[-16, +16]$~\cite{Kersting_2018}. 
The resulting effective regulator ratio ranges from $r_{\min} = 0.9$ to $r_{\max} =1.1$. In the present paper, the effective regulator ratio is assumed to be continuous in the interval $[r_{\min}, r_{\max}]$~\cite{Hafez_OPF_SVR, Robbins2016}. In addition, the SVR is non-gang-operated, that is, the effective regulator ratio can be chosen independently for each phase. The inverse relation between the tap position $t^{\phi}_{nn^{\prime}}$ and effective regulator ratio $r^{\phi}_{nn^{\prime}}$ is given by
\begin{align}
    t^{\phi}_{nn^{\prime}} = \mathrm{round}\bigg[\frac{1-r^{\phi}_{nn^{\prime}}}{0.00625}\bigg] \label{eq:ratio2tap}
\end{align}

In case of type-A SVR, voltage relationship~\eqref{eq:voltage gain} is replaced by
\begin{equation}
    {v}_{n^{\prime}} = {A}_{n^{\prime}n}{v}_{n} 
\end{equation}
and the gain matrix remains the same as~\eqref{eq:gain matrix} with effective regulator ratio given by $r^{\phi}_{nn^{\prime}} = 1 + 0.00625 t^{\phi}_{nn^{\prime}}$. The current relationship~\eqref{eq:current gain} and the ensuing power flow equations and relaxations must be adjusted accordingly. In the remainder of the development, we focus on type-B SVR for brevity.   

\subsection{Branch Flow Equations}
To formulate the power balance at bus $m$, consider the edges $n \rightarrow m \rightarrow k$, and multiply~\eqref{eq:ohmslaw} by $\bar{i}_{nm}$:
\begin{align}
    v_{n}\bar{i}_{nm} &= v_{m}\bar{i}_{nm}+Z_{nm}i_{nm}\bar{i}_{nm} 
    \label{eq:powerbalance1} \end{align}
Kirchoff's current law (KCL) at node $m$ is  stated as follows:
    \begin{align}
    \bar{i}_{nm}+\bar{i}_{m} &= \sum_{(m,k) \in \Eb} \bar{i}_{mk} \label{eq:KCL}
    \end{align}
Specifically, ${i}_{m}\in \mathbb{C}^{3}$ is the net current injection at bus $m \in \Nb_{+}$, which is generically a sum of currents from constant-power sources with net complex power ${s}^{\mathrm{c}}_{m}\in \mathbb{C}^{3}$ and from constant-admittance elements, such as shunt capacitor banks, with admittance $Y_{m}$ connected to bus $m$. The net current injection at bus $m$ is thus given by 
\begin{align}
   {i}_{m}& = \diag(v^{*}_{m})^{-1}({s}^{\mathrm{c}}_{m})^{*}-Y_{m}v_{m} \label{eq:netcurrentinjection}
\end{align}

Substituting~\eqref{eq:KCL} in~\eqref{eq:powerbalance1} yields
  \begin{align}
   v_{n}\bar{i}_{nm} &= v_{m} \left[\sum_{(m,k) \in \Eb} \bar{i}_{mk} - \bar{i}_{m}\right]+Z_{nm}i_{nm}\bar{i}_{nm}, \notag \\
   & \mspace{250mu} (n,m) \in \Eb_{\Tb} \label{eq:powerbalance2}
\end{align}
Taking conjugate transpose of~\eqref{eq:netcurrentinjection}, substituting in~\eqref{eq:powerbalance2}, and then taking $\diag$ yields
\begin{align}
   \diag(v_{n}\bar{i}_{nm}) &= \sum_{(m,k) \in \Eb} \diag(v_{m}\bar{i}_{mk})-s^{\mathrm{c}}_{m}\notag \\
   &+\diag(v_{m}\bar{v}_{m}\bar{Y}_{m}) 
   + \diag(Z_{nm}i_{nm}\bar{i}_{nm}),\notag \\
  & \mspace{250mu} (n,m) \in \Eb_{\Tb} \label{eq:powerbalance3}
\end{align}
The previous represents the power flow equations for any branch $(n,m) \in \Eb_{\Tb}$.

Attention is turned next to edges $(n,n')\in \Eb_{\Rb}$. Specifically, the power balance at the secondary $n'$ of an SVR is derived upon taking conjugate transpose of~\eqref{eq:current gain}, multiplying from the left with $v_n$, and invoking~\eqref{eq:voltage gain} to obtain 
\begin{align}
    \diag(v_{n}\bar{i}_{nn^{\prime}}) & =\diag(v_{n^{\prime}} \bar{i}_{n^{\prime}m}), \  (n,  n^{\prime}) \in \Eb_{\Rb}, (n^{\prime}, m) \in \Eb_{\Tb}  \label{eq:KCLatSVRnode}
\end{align}
It is worth noting that~\eqref{eq:KCLatSVRnode} holds true for grounded-wye grounded-wye SVRs because the gain matrix ${A}_{nn^{\prime}}$ in~\eqref{eq:gain matrix} is diagonal~\cite{Hafez_OPF_SVR, Alsaleh_2021, Robbins2016}. \\

\subsection{Branch Flow Optimal Tap Selection}
The objective is to minimize the real power import from the substation given by
\begin{align}
{C}_{} & = \mathrm{Re}\left\{\sum_{(\Sb,m)\in\Eb} \bm{1}^{\top}_{3} \mathrm{diag}(v_{\Sb}\bar{i}_{{\Sb}, m})\right\} \label{eq:powerimport}
\end{align}
where $\bm{1}_{3}$ is a $3 \times 1 $ vector of all ones. 
The branch-flow optimal tap selection \textit{(BF-OPTS)} problem $(\text{P1})$ is stated next:
\begin{subequations}
\label{eq:b-opf-all}
\begin{align}
\text{(P1)}
~~\min~~~& {C}_{} \\
\text{over}~~~&\mspace{-10mu} \{v_{m}\}_{m \in \Nb_{+}}, \{i_{nm}\}_{(n,m)\in \Eb_{\Tb}}, \notag \\
&\mspace{-30mu} \{A_{nn^{\prime}}\}_{(n, n^{\prime}) \in \Eb_{\Rb}} , \{i_{nn^{\prime}}\}_{(n,n^{\prime})\in \Eb_{\Rb}}, \{r^{\phi}_{nn^{\prime}}\}_{(n,n^{\prime})\in \Eb_{\Rb}}  \\
&\eqref{eq:ohmslaw}, \eqref{eq:voltage gain}, \eqref{eq:current gain}, \eqref{eq:gain matrix}, \eqref{eq:powerbalance3}, \eqref{eq:KCLatSVRnode}, \eqref{eq:powerimport} \\
&v_{\min} \leq |v_{m}| \leq v_{\max} \label{eq:voltagecons} \\
&r_{\min} \leq r^{\phi}_{nn^{\prime}} \leq r^{\max},\ (n, n^{\prime}) \in \Eb_{\Rb} \label{eq:tapcons}
\end{align}
\end{subequations}
The \textit{BF-OPTS} problem $(\text{P1})$ is nonconvex and hard to solve.  The nonconvexity stems from the bilinear equalities in~\eqref{eq:voltage gain} and~\eqref{eq:current gain}; quadratic equalities in~\eqref{eq:powerbalance3} and~\eqref{eq:KCLatSVRnode}; and the left-hand side of~\eqref{eq:voltagecons}. 

The ensuing section develops a linear approximation to (P1). The nonconvexities arising from the power flow equations are alleviated using the \textit{LinDist3Flow} model. In addition, the nonconvexities stemming from the SVR model are relaxed to linear constraints. These manipulations are presented next. 

\section{\text{LinDist3Flow}-based Optimal Tap Selection}
First, the following auxiliary matrix variables are introduced: $V_{n} = v_{n} \bar{v}_{n}$, 
$n \in \Nb$; $S_{nm} = v_{n}\bar{i}_{nm}$ and $I_{nm} = i_{nm} \bar{i}_{nm}$,  $(n,m) \in \Eb$; $S_{nn^{\prime}} = v_{n}\bar{i}_{nn^{\prime}}$ and $I_{nn^{\prime}} = i_{nn^{\prime}} \bar{i}_{nn^{\prime}}$,  $(n,n^{\prime}) \in \Eb_{\Rb}$.

Upon multiplying both sides of~\eqref{eq:ohmslaw} by their conjugate transposes $\bar{(.)}$, i.e., $\bar{v}_{n}$ on the left and $(\bar{v}_{m} +\bar{i}_{nm}\bar{Z}_{nm})$ on the right,~\eqref{eq:ohmslaw} can be written  as 
\begin{align}
 V_{n} = V_{m}+2 \ \mathrm{Re}\{v_{m}\bar{i}_{nm}\bar{Z}_{nm}\}+\underbrace{Z_{nm}I_{nm}\bar{Z}_{nm}}_{{H}_{nm} = \text{higher order term}} \label{eq:conjugateohmslaw} 
\end{align}
The higher order term $H_{nm}$ in~\eqref{eq:conjugateohmslaw}  represents the change in voltage associated with loss. 

Define next the auxiliary vector variables $\tilde{v}_{n} = \diag(V_{n}) \in \mathbb{R}^{3}$, $ n \in \Nb$, as the vectors of squared voltage magnitudes, $\tilde{S}_{nm} = \diag(S_{nm}) \in \mathbb{C}^{3}$, and $\tilde{H}_{nm} = \diag(H_{nm}) \in \mathbb{R}^{3}$. 
Define further the complex line current $i^{\phi}_{nm} = (\tilde{S}^{\phi}_{nm}/v^{\phi}_{m})^{*}$, $(n,m) \in \Eb$, $\phi \in \{a, b,c\}$ and rotation matrix $\Gamma_{m} = \begin{bmatrix} 1 & \gamma^{ab}_{m} & \gamma^{ac}_{m}\\
\gamma^{ba}_{m} & 1 & \gamma^{bc}_{m} \\
\gamma^{ca}_{m} & \gamma^{cb}_{m} & 1 \end{bmatrix}$ with entries given by $\gamma^{\phi \psi}_{m} = \frac{v^{\phi}_{m}}{v^{\psi}_{m}}$, $\phi,\psi\in\{a,b,c\}$, and $\phi \neq \psi$. Following~\cite{Arnold_2016} and invoking the previous definitions of the complex current and rotation matrix, it follows upon taking $\mathrm{diag}$ of~\eqref{eq:conjugateohmslaw} that
\begin{align}
  \mspace{-10mu}  \tilde{v}_{n} & = \tilde{v}_{m} + 2\ \mathrm{Re}\{(\Gamma_{m} \odot \bar{Z}_{nm}) \tilde{S}_{nm} \}+\tilde{H}_{nm}, (n,m) \in \Eb_{\Tb} \label{eq:voltagerotationvector}
\end{align}
where the operator $\odot$ denotes element-wise product. 

Using the definitions of $\tilde{S}_{nm}$, $\tilde{v}_{m}$, and $I_{nm}$, eq.~\eqref{eq:powerbalance3} is written as
\begin{align}
    \tilde{S}_{nm} &= \mspace{-10mu}\sum_{(m,k) \in \Eb} \tilde{S}_{mk}-s^{\mathrm{c}}_{m}+\bar{Y}_{m}\tilde{v}_{m}
   + \underbrace{\diag(Z_{nm}I_{nm})}_{\tilde{L}_{nm}=\text{higher order loss term}} , \notag \\
 &\mspace{-50mu}   (n,m) \in \Eb_{\Tb}, \ m \in \Nb_{+} \label{eq:powerbalance4}
\end{align}
where $\tilde{L}_{nm} \in \mathbb{C}^{3}$ denotes the higher order loss term. 

Note that~\eqref{eq:voltagerotationvector} and~\eqref{eq:powerbalance4} are still nonlinear. To derive linear approximations,  two assumptions are adopted~[\citenum{sankur2016linearized}, Section.~II]:  (1) The rotation matrix entries $\gamma^{\phi \psi}_{m}$ are constant; and (2) the higher order terms in~\eqref{eq:voltagerotationvector} and~\eqref{eq:powerbalance4} are constant.
With the two aforementioned assumptions,~\eqref{eq:voltagerotationvector} and~\eqref{eq:powerbalance4} become linear, and the latter is written as
  \begin{align}
    \tilde{S}_{nm} & = \mspace{-10mu}\sum_{(m,k) \in \Eb_{\Tb}} \tilde{S}_{mk}-s^{\mathrm{c}}_{m}+\bar{Y}_{m}\tilde{v}_{m}+\tilde{L}_{mn}, (n,m) \in \Eb_{\Tb} 
    \label{eq:powerbalance5}
\end{align}
where $\Gamma_{m}$, $\tilde{H}_{nm}$, and $\tilde{L}_{nm}$ are constants in~\eqref{eq:voltagerotationvector} and~\eqref{eq:powerbalance5}. 

It is worth pointing out that the aforementioned assumptions are not overly restrictive, as there are different ways to select the values of constants $\Gamma_{m}$, $\tilde{H}_{nm}$, and $\tilde{L}_{mn}$ to improve the quality of the approximation. Specifically, these can be computed from an initial power flow solution with a specific SVR tap setting (e.g., taps set to zero) and kept constant afterwards.
An even simpler approach is to set the higher order terms $\tilde{H}_{nm}$ and $\tilde{L}_{nm}$ to zero and assume that voltages are approximately balanced (i.e., approximately equal in magnitude and $120^{\circ}$ apart), which yields $\gamma^{ab}_{m} = \gamma^{bc}_{m} = \gamma^{ca}_{m} \approx \alpha$ and $\gamma^{ac}_{m} = \gamma^{ba}_{m} = \gamma^{cb}_{m} \approx \alpha^{2}$, with $\alpha = 1\angle 120^{\circ}$~\cite{Arnold_2016, gan2014convex}. Another approach is proposed in~\cite{RobbinsOPF}, which considers a linear approximation to the higher order terms (rather than treating them as constant). The rotation matrix and higher order terms may also be re-computed based on the measured power flows by solving the $\textit{LinDist3Flow}$ iteratively in a successive approximation fashion. The previously mentioned references utilize $\textit{LinDist3Flow}$ to solve the OPF problem without optimal tap selection. The selection of higher order terms for solving the optimal tap selection problem with $\textit{LinDist3Flow}$ is explained in Section~IV.

Eq.~\eqref{eq:voltagerotationvector} and~\eqref{eq:powerbalance5} correspond to the \emph{LinDist3Flow}-approximated voltage and power balance equations for non-SVR edges. The respective constraints for SVR edges are derived next.
Following~\cite{Robbins2016},~\eqref{eq:voltage gain} is relaxed with valid inequalities as follows:
\begin{align}
    (r_{\min})^2  \tilde{v}_{n^{\prime}} \leq \tilde{v}_{n} \leq (r_{\max})^2 \tilde{v}_{n^{\prime}}, (n, n^{\prime}) \in \Eb_{\Rb},\label{eq:voltagegainrelaxation} 
\end{align}
Using the auxiliary variables $\tilde{S}_{n^{\prime}m}$ and $\tilde{S}_{nn^{\prime}}$, the SVR power balance~\eqref{eq:KCLatSVRnode} is stated below
\begin{align}
  \tilde{S}_{nn^{\prime}} = \tilde{S}_{n^{\prime}m}, \  (n,  n^{\prime}) \in \Eb_{\Rb}, (n^{\prime}, m) \in \Eb_{\Tb}  \label{eq:SVRpowerbalance}   
\end{align}
It is worth noting that while the power balance in~\eqref{eq:powerbalance5} is approximate (since losses are ignored or treated as constant), the power balance in~\eqref{eq:SVRpowerbalance} is exact. 

The objective function is written in terms of the new optimization variables as
\begin{equation}
\label{eq:approxobj}
    \tilde{C}=\mathrm{Re}\left\{ \sum_{(\Sb,m)\in\Eb}\bm{1}^{\top}_{3} \tilde{S}_{\Sb,m}\right\}
\end{equation}
It should be noted that the objective in~\eqref{eq:approxobj} is an approximation to~\eqref{eq:powerimport}, i.e., $\tilde{C} \approx C$.
Specifically,~\eqref{eq:powerimport} and~\eqref{eq:approxobj} would be equal if the higher order loss terms and the entries of rotation matrix were computed from the optimal power flow solution, in which case the relationship $\tilde{S}_{\Sb,m} = \diag(S_{\Sb, m}) = \diag(v_{\Sb} \bar{i}_{S,m})$ would be exact. 

The linear approximation $\textit{LinDist3Flow-OPTS}$ to the nonconvex optimal tap selection problem $\text{(P1)}$ is stated next:
\begin{subequations}
\label{eq:LinDist3Flow-opf-all}
\begin{align}
\text{(P2)}
~~\min~~~& \tilde{C}_{} \\
\text{over}~~~&\{\tilde{v}_{m}\}_{m \in \Nb_{+}}, \{\tilde{S}_{nm}\}_{(n,m)\in \Eb_{\Tb}}, \{\tilde{S}_{nn^{\prime}}\}_{(n,n^{\prime})\in \Eb_{\Rb}} \\
&~\eqref{eq:voltagerotationvector}, \eqref{eq:powerbalance5}, \eqref{eq:voltagegainrelaxation}, \eqref{eq:SVRpowerbalance} \\
&(v_{\min})^2 \leq \tilde{v}_{m} \leq (v_{\max})^2 \label{eq:voltagelimits} 
\end{align}
\end{subequations}
It should be noted that the $\textit{LinDist3Flow-OPTS}$ problem $\text{(P2)}$ is convex. Specifically, it is a linear programming (LP) problem  which can be easily solved using off-the-shelf solvers. The performance of $\textit{LinDist3Flow-OPTS}$ is assessed and compared to SDP based formulations in the next section. 

\section{Numerical Results}
This section evaluates the performance of~$\textit{LinDist3Flow-OPTS}$ in comparison  to conventional nonlinear and SDP-based approaches for optimal tap selection. 

The standard IEEE 13-bus, 123-bus, and 8500-node networks with a variety of three-, two-, and one-phase lines are used. Transformers and SVRs are modeled with grounded-wye connections. Short lines take the place of switches. Line shunt admittances are ignored, but capacitors are accounted for in accordance with the documentation. All loads are converted to wye-connected constant-power loads. The slack bus voltage $v_{\Sb}$ is set to $v_{\Sb} = |v_{\Sb}| \times \{1, e^{-j 2\frac{\pi}{3}}, e^{j 2\frac{\pi}{3}}\}$, where the value of $|v_{\Sb}|$ is respectively set to 1.0 p.u. for the IEEE 13- and 123-bus DNs and 1.05 p.u. for the IEEE 8500-node DN~\cite{Hafez_OPF_SVR}. 
The $\textit{LinDist3Flow-OPTS}$ is solved using the MATLAB-based toolbox YALMIP~\cite{Lofberg2004} with MOSEK as optimization solver~\cite{mosek}.  

The performance of the $\textit{LinDist3Flow-OPTS}$ is evaluated with respect to  three previously available SDP formulations which are termed as $\textit{CI-OPTS}$~(full SDP relaxation of admittance-based $\textit{OPTS}$) \cite{Robbins2016}, $\textit{CG-OPTS}$~(chordal SDP relaxation of admittance-based $\textit{OPTS}$)~\cite{Liu_2018}, and $\textit{MB-OPTS}$~(tight relaxation of branch flow-based $\textit{OPTS}$)~\cite{Hafez_OPF_SVR} along with a traditional NLP formulation $\textit{BF-OPTS}$ in~\eqref{eq:b-opf-all}. The codes available at~\url[https://github.com/hafezbazrafshan/BranchFlowMultiphaseVRs] are used to solve SDP-relaxations $\textit{CI-OPTS}$, $\textit{CG-OPTS}$, and $\textit{MB-OPTS}$ and NLP formulation $\textit{BF-OPTS}$ respectively. The SDP relaxations are solved in CVX~\cite{Grant2014} using MOSEK with the exception of $\textit{CI-OPTS}$ for the IEEE-123 bus DN which is solved with solver SDPT3 in CVX, while $\textit{BF-OPTS}$ is solved using YALMIP with solver IPOPT.  
All simulations are run on a 2.60-GHz, intel core i7 computer with 16 GB of RAM. 

It should be noted that the effective regulator ratio is not an explicit variable in $\textit{LinDist3Flow-OPTS}$ problem $\text{(P2)}$. Therefore, after solving the $\textit{LinDist3Flow-OPTS}$ problem, the effective regulator ratio is retrieved using the relation $r^{\phi}_{nn^{\prime}} = \sqrt{\frac{\tilde{v}^{\phi}_{n^{}}}{\tilde{v}^{\phi}_{n^{\prime}}}} \ (n, n^{\prime}) \in \Eb_{\Rb}$ and the SVR taps are then computed using~\eqref{eq:ratio2tap}. Furthermore, since $\textit{LinDist3Flow-OPTS}$ is an approximation of actual nonlinear power flows, upon fixing the effective regulator ratios, the Z-Bus method is run to obtain actual voltage solutions~\cite{Zbushafez,ConvZBus}. However, other methods for obtaining voltages, such as the forward-backward sweep, may also be used~\cite{Kersting_2018}.

\begin{table}[t]
\centering
\tiny
\caption{Maximum absolute p.u. voltage magnitude difference: $\textit{LinDist3Flow}$ vs. Z-bus power flow}
\begin{center}
 \begin{tabular}{|p{2.0cm}|ccccc|} \hline
DN & phase a  & phase b & phase c & $\min{\check{v}}$& $\min{{v}}$ \\
 \hline\hline
IEEE 13-bus & 0.009 & 0.007 & 0.01 & 0.88 & 0.89 \\
IEEE 123-bus & 0.02 & 0.008 & 0.008 & 0.86 & 0.88\\
IEEE 8500-node & 0.06 & 0.04 & 0.008& 0.84 & 0.90\\
  \hline
\end{tabular}
\label{tab:p.u.diff}
\end{center}
\end{table}

Prior to solving $\textit{LinDist3Flow-OPTS}$, the maximum absolute p.u. differences between voltage magnitudes computed by the $\textit{LinDist3Flow}$ power flow and Z-bus power flow are evaluated by setting the SVR taps and higer order terms $\tilde{H}_{nm}$, and $\tilde{L}_{mn}$ to zero and reported in Columns 2--5 of Table~\ref{tab:p.u.diff}.  The table indicates that for the IEEE 8500-node DN, the maximum absolute p.u. difference of voltage magnitudes varies significantly across phases. This is because the IEEE 8500-node DN has $1543$ single-phase buses, the majority of which are on phases $a$ and $b$.
Furthermore, Columns 4 and 5 of Table~\ref{tab:p.u.diff} report the minimum voltage magnitudes computed from the Z-bus and \textit{LinDist3Flow} power flow methods defined as $\min{\check{v}} = \underset{n, \phi}{\min}~ |\check{v}^{\phi}_{n}|$ and $\min{{v}} = \underset{n, \phi}{\min}~ |{v}^{\phi}_{n}|$, where  $\check{v}_n^{\phi}$ and ${v}_n^{\phi}$ are the voltage profiles obtained from the Z-bus method and $\textit{LinDist3Flow}$ respectively. 

Table~\ref{tab:p.u.diff} suggests that for the IEEE 13-bus, 123-bus, and 8500-node DNs,  $\textit{LinDist3Flow}$ somewhat overestimates the voltage magnitudes, i.e, $\min{{v}} > \min{\check{v}}$. This is due to the fact that the higher order terms $\tilde{H}_{nm}$ and $\tilde{L}_{nm}$ in~\eqref{eq:conjugateohmslaw} and~\eqref{eq:powerbalance5} have been set to zero. Therefore, the voltage magnitudes obtained by solving an OPF utilizing $\textit{LinDist3Flow}$ such as $\text{(P2)}$  while ignoring higher order terms may turn out to be infeasible when the nonlinear power flows are computed based on the optimized effective regulator ratios. To circumvent this issue, the following approaches can be used: (1) heuristically adjust the minimum voltage limit $v_{\min}$ for heavily loaded DNs to tackle under-voltage issues (respectively, $v_{\max}$ in lightly loaded DNs for over-voltage issues)~\cite{Stuhlmacher_2020}; and (2) consider nonzero higher order terms~\cite{RobbinsOPF}. The specific choices for $v_{\min}$ and the higher order terms adopted in the present paper for $\textit{LinDist3Flow-OPTS}$ are detailed next. 

Based on the previously mentioned observations, the $\textit{LinDist3Flow-OPTS}$ problem is solved upon initializing the higher order terms $\tilde{H}_{nm}$ and $\tilde{L}_{nm}$ and the entries of $\Gamma_m$ based on the power flow computed by the Z-bus method with the SVR taps set to zero.
Parameters $v_{\min}$ and $v_{\max}$ are respectively set to 0.9 p.u. and 1.10 p.u. for all SDP-based relaxations ($\textit{CI-OPTS}$, $\textit{CG-OPTS}$, and $\textit{MB-OPTS}$) and NLP formulation $\textit{BF-OPTS}$ in the IEEE 13-bus DN. They are set to 0.93 p.u. and 1.10 p.u. for $\textit{LinDist3Flow-OPTS}$ in the IEEE 13-bus DN. 
Parameters $v_{\min}$ and $v_{\max}$ are respectively set to 0.9 p.u. and 1.10 p.u. for all SDP-based relaxations, $\textit{BF-OPTS}$, and  $\textit{LinDist3Flow-OPTS}$, in the IEEE 123-bus and IEEE 8500-node DNs.

\begin{table*}[t]
\tiny
\centering
\caption{Comparisons between SDP-relaxations, $\textit{BF-OPTS}$, and $\textit{LinDist3Flow-OPTS}$ }
\begin{center}
 \begin{tabular}{|p{1.5cm}|ccccccccc|} \hline
DN & Method & $\hat{C}$ & $\check{C}$ & min ($\check{v}$) & max ($\check{v}$) & $\mathrm{feas.}$& $\check{v}_{\mathrm{unb}}$ & Time (sec) &  Gap ($\%$)   \\ 
 \hline 
13-bus & $\textit{CI-OPTS}$ & 0.7121 & 0.7135 & 0.99 & 1.12 &$\mathrm{infeas.}$ & 6.54 & 1.78 &--\\
            & $\textit{CG-OPTS}$ & 0.7141 & 0.7141 &  0.99 & 1.10 & $\mathrm{feas.}$ & 6.61 & 1.91& 0 \\
            & $\textit{MB-OPTS}$ & 0.7135 & 0.7135 &  0.99 & 1.10 & $\mathrm{feas.}$ & 5.55 & 2.09& 0\\
            & $\textit{BF-OPTS}$ & -- & 0.7135 & 0.99 & 1.10 & $\mathrm{feas.}$& 5.68 & 0.4& 0\\
            & $\textit{LinDist3Flow-OPTS}$ & --&  0.7176 &0.92 & 1.04 & $\mathrm{feas.}$ & 4.15 & 0.16 & 0.5\ $(\textit{CI-OPTS})$, 0.4\ $(\textit{CG-OPTS})$, 0.5\ $(\textit{MB-OPTS})$ \\
            & $\textit{No SVR}$ &-- & 0.7198 & 0.88 & 1.00 & $\mathrm{infeas.}$& 6.5 & --&--\\ 
            \hline \hline 
123-bus & $\textit{CI-OPTS}$ & 0.7215 & 0.7219 & 0.96 & 1.11 &$\mathrm{infeas.}$& 4.69 & 2.85 & --\\
            & $\textit{CG-OPTS}$ & 0.7222 & 0.7222 &  0.95 & 1.10 & $\mathrm{feas.}$ & 4.55 & 2.7& 0 \\
            & $\textit{MB-OPTS}$ & 0.7218 & 0.7218 &  0.96 & 1.09 & $\mathrm{feas.}$& 3.09 & 3.07& 0\\
            & $\textit{BF-OPTS}$ &  --   & 0.7218 & 0.96 & 1.10 & $\mathrm{feas.}$ & 3.05 & 3.0& 0\\
            & $\textit{LinDist3Flow-OPTS}$ & --&  0.7246 &0.90 & 1.02 & $\mathrm{feas.}$ & 3.75 & 0.3 & 0.4\ $(\textit{CI-OPTS})$, 0.3\ $(\textit{CG-OPTS})$, 0.4\ $(\textit{MB-OPTS})$ \\
            & $\textit{No SVR}$ &-- & 0.7283 & 0.86 & 1.00 & $\mathrm{infeas.}$& 8.1 & --&--\\ 
            \hline \hline 
8500-node & $\textit{CI-OPTS}$ & 0.4129 & 0.4180 & 0.96 & 1.13 & $\mathrm{infeas.}$& 7.64 & 15 &--\\
            & $\textit{CG-OPTS}$ & 0.4184 & 0.4176 &  0.98 & 1.10 & $\mathrm{feas.}$ & 6.2 & 16& 0.55 \\
            & $\textit{MB-OPTS}$ & 0.4161 & 0.4181 &  0.93 & 1.09& $\mathrm{feas.}$ & 6.00 & 17& 0.47\\
            & $\textit{BF-OPTS}$ &   -- & 0.4166 & 0.98 & 1.10& $\mathrm{feas.}$ & 3.0& 800& 0.12\\
            & $\textit{LinDist3Flow-OPTS}$ & --&  0.4206 &0.94 & 1.07 & $\mathrm{feas.}$& 7.95 & 0.5 & 1.8\ $(\textit{CI-OPTS})$, 0.5\ $(\textit{CG-OPTS})$, 1.08\ $(\textit{MB-OPTS})$ \\
            & $\textit{No SVR}$ &-- & 0.4252 & 0.84 & 1.05 & $\mathrm{infeas.}$& 12.3 & --&--\\ 
            \hline 
\end{tabular}
\label{tab:comparison}
\end{center}
\end{table*}

The results of the optimization are summarized in Table~\ref{tab:comparison}. Formulations $\textit{CI-OPTS}$, $\textit{CG-OPTS}$, and $\textit{MB-OPTS}$ are relaxations of the optimal tap selection problem and thus provide lower bounds for the optimal value of (P1). The resulting optimal values are listed in Column 3 of Table~\ref{tab:comparison} under the notation $\hat{C}$. No $\hat{C}$ value is reported for $\textit{BF-OPTS}$ or $\textit{LinDist3Flow-OPTS}$, because these do not provide a lower bound on the objective. The last row for each network (\textit{No SVR}) amounts to power flow with taps set to zero.  
While $\textit{BF-OPTS}$ and $\textit{MB-OPTS}$ explicitly include the effective regulator ratio $r_{nn'}^{\phi}$ as optimization variables, problems $\textit{CI-OPTS}$, $\textit{CG-OPTS}$, and $\textit{LinDist3Flow-OPTS}$ do not, and the effective regulator ratio must be computed from the voltage variables of the corresponding formulation, using the expression given earlier in this section. For each of problems $\textit{CI-OPTS}$, $\textit{CG-OPTS}$, $\textit{MB-OPTS}$, $\textit{BF-OPTS}$, and $\textit{LinDist3Flow-OPTS}$, the resulting effective regulator ratios $r_{nn'}^{\phi}$ are used as inputs for the Z-Bus method to produce voltage profiles denoted by $\check{v}$ that conform to the nonlinear power flow equations.

Columns 4--10 provide the results computed based on $\check{v}$.
The actual objective value $\check{C}$ reported in Column 4 is computed from the power flow solution $\check{v}$ as follows
\begin{equation}
\label{eq:zbuspowerimport}
   \check{C} = \mathrm{Re}\left\{\bm{1}^{\top}_{3} \mathrm{diag}(v_{\Sb}\check{\bar{v}}\mathbf{\bar{Y}_{\Sb}})\right\}
\end{equation}
where $\mathbf{Y}_{\Sb}$ is the bus admittance matrix from the set $\Nb$ to the slack bus and the entries of vector $\check{v}$ are the phase voltages at all buses, including the slack bus. The power import objectives in~\eqref{eq:powerimport} and~\eqref{eq:zbuspowerimport} are equivalent, and the proof is provided in the Appendix.

The minimum and maximum magnitudes of actual power flow voltages are respectively given by $\min{\check{v}} = \underset{n, \phi}{\min}~ |\check{v}^{\phi}_{n}|$ and $\max{\check{v}} = \underset{n, \phi}{\max}~|\check{v}^{\phi}_{n}|$ and are listed in Columns 5 and 6. Column 7 reports whether the resulting voltage profile $\check{v}$ is feasible, i.e., within the bounds $v_{\min}$ and $v_{\max}$. The voltage unbalance $\check{v}_{\mathrm{unb}}$ defined by ANSI~[\citenum{Kersting_2018}, eq. (7.1)] is provided in Column 8. Computation times are listed in Column 9. 

The quantity $\mathrm{Gap}$ in Column 10 pertains to the optimality gap computed based on a feasible objective value $\check{C}$ and a lower bound $\hat{C}$ and defined by $\mathrm{Gap}=\frac{\check{C}-\hat{C}}{\hat{C}}\times 100\%$. No optimality gap is reported for the IEEE 13-bus, IEEE123-bus, and 8500-node networks under $\textit{CI-OPTS}$ because the overall method did not return feasible voltages for these networks (cf. Columns 5--7). The optimality gaps for $\textit{CG-OPTS}$ and $\textit{MB-OPTS}$ are computed based on the lower bound provided by the respective relaxation, with the exception of $\textit{CG-OPTS}$ for the IEEE 8500-node DN, where the lower bound provided by $\textit{MB-OPTS}$ is used (we observe that $\hat{C}>\check{C}$ for $\textit{CG-OPTS}$ in Table~\ref{tab:comparison} due to apparent numerical issues). The optimality gap for $\textit{BF-OPTS}$ is computed based on the lower bound from $\textit{MB-OPTS}$. For completeness, the optimality gap of $\textit{LinDist3Flow-OPTS}$ is reported against all three available SDP relaxations. The optimality gap is an estimate of how close a feasible solution is to the optimal value, and it is thus reasonable to choose the tightest lower bound to the optimal value available to draw conclusions. 

The following key observations are made from Table~\ref{tab:comparison}:
\begin{enumerate}
    \item For the IEEE 13- and 123-bus DNs, the $\textit{LinDistFlow-OPTS}$ yields an optimal value with optimality gap is $0.5\%$ or less with respect to all three SDP-based relaxations.
    The optimality gap for the IEEE 8500-node network is $1.08\%$ when benchmarked against the lower bound provided by $\textit{MB-OPTS}$. 
    Overall, the reported optimality gaps are quite small.
    \item Column $8$ reveals that solving $\textit{LinDistFlow-OPTS}$ is significantly faster than any of the other methods. Specifically, for the IEEE 8500-node DN,  $\textit{LinDistFlow-OPTS}$ achieves a remarkable speedup of at
least 30 times when compared to $\textit{BF-OPTS}$.  
\item Although SDP formulations can outperform $\textit{LinDist3Flow-OPTS}$ in terms of providing smaller cost, the number of variables for such formulations grows significantly faster with the number of buses than their LP counterpart, in addition to the complexity of accommodating the positive semidefinite cone constraints. Comparable results can be achieved with $\textit{LinDist3Flow-OPTS}$, while requiring less computational effort.
\item The $\textit{LinDist3Flow-OPTS}$ yields feasible voltage profiles for all test networks verified with the nonlinear power flow solver provided by the Z-Bus method.  
\item The $\textit{LinDist3Flow-OPTS}$ achieves the lowest voltage unbalance for the IEEE 13-bus DN when compared to other formulations, and similar voltage unbalance to the other methods for the IEEE 123-bus DN. The voltage unbalance achieved by $\textit{LinDist3Flow-OPTS}$ in the IEEE 8500-node DN is only lightly higher than one achieved by SDP formulations. It is worth noting that no specific objective is included here to encourage the minimization of voltage unbalance explicitly. 
\end{enumerate}

Finally, Table~\ref{tab:Taps} provides the optimal taps provided by the different formulations. Although the various methods yield different tap selections, the quality of the solution for each method can be better assessed from the results listed in Table~\ref{tab:comparison}.

\begin{table}[t]
\centering
\tiny
\caption{Optimal Taps Obtained by Various Formulations}
\begin{center}
 \begin{tabular}{|p{1.0cm}|ccccc|} \hline
SVR ID & \textit{CI-OPTS}  & \textit{CG-OPTS}  & \textit{MB-OPTS}& \textit{BF-OPTS} & \textit{LinDist3Flow-OPTS}\\
 \hline\hline
 13-1 & 15, 15, 15 & 13, 13, 13 & 15, 13, 15 & 15, 13, 15 & 4, -8, 7 \\
 \hline \hline
 123-1& 10, 5, 7 & 3, 3, 3 & 11, 3, 7 & 11, 3, 7 & 4, 1, 2 \\
 123-$2^{*}$ & 9, 4 & 7, 7 & 11, 7 & 11, 7 & 3, 1 \\
 123-$3^{\dagger}$ & 7 & 9 & 9 & 9 & -15 \\
 123-4 & 16, 16, 16 & 16, 16, 16 & 16, 16, 16 & 16, 16, 16& 10, 5, 8 \\
 \hline \hline
 8500-1 & 9, 9, 8 & 9, 9, 7 & 6, 7, 8 & 9, 9, 8 & 5, 4, -3 \\
 8500-2& 3, 5,4 & 4, 3, -1 & 4, 3, 1 & 6, 4, 1& 7, 5, 1 \\
 8500-3& 7, 6, 3 & 7, 6, 2 & 12, 10, 2 & 14, 9, 2 & 9, 5, -1 \\
 8500-4& 2, 5, 1 & 3, 3, -1 & 0, -1, -14& 10, 11, -8 & 4, 7, 3\\
 \hline
\end{tabular}
\label{tab:Taps}
\end{center}
$*$ SVR ID $2$ is two-phase wye \
$\dagger$ SVR ID $3$ is single-phase wye.
\end{table}

\section{Conclusions and Future Work}
This paper analyzes the performance of $\textit{LinDist3Flow}$ for optimal tap selection of wye-connected SVRs. The numerical results carried out on standard IEEE test distribution feeders reveal  that the $\textit{LinDist3Flow-OPTS}$ performs reasonably well when compared with existing SDP-based and nonlinear approaches with an optimality gap of \emph{approximately} 1\% or less. The chief advantage of  $\textit{LinDist3Flow-OPTS}$ is the  significant reduction in  computational effort. Future work will incorporate other types of SVRs, PV inverter dispatch, as well as SVR placement problems that are critical for enhancing PV hosting capacity in unbalanced DNs. Another fruitful direction is to evaluate the performance of $\textit{LinDist3Flow}$ with respect to other linearizations such as first-order Taylor series approximation~\cite{Garces_2016}, fixed-point linearization~\cite{Bernstein_2017}, generalized $\textit{LinDist3Flow}$~\cite{Huang_2021}, and forward-backward sweep methods~\cite{Girigoudar_2021} in the context of optimal tap selection. It is worth pointing out that for each of the aforementioned linearizations, it is a research issue in its own right to include the optimal tap selection owing to the nonlinear relationships between the constituent SVR variables. Future work will also focus on evaluating the performance of $\textit{LinDist3Flow-OPTS}$ by solving it iteratively in a successive approximation fashion upon updating the higher order terms.

The $\textit{LinDist3Flow-OPTS}$ in the present paper implements optimal tap selection in an $\emph{open-loop}$ fashion similar to that of~\cite{Robbins2016, Li_2018, Hafez_OPF_SVR, Alsaleh_2021}. That is, the $\textit{LinDist3Flow-OPTS}$ takes into account the net nodal injections and produces taps to limit the voltages to respect~\eqref{eq:voltagelimits}.
It is also possible to formulate the problem in a \emph{closed-loop} fashion, whereby the secondary-side voltage is used as input for tap selection, or even a load center elsewhere in the network~\cite{Kersting_2018}, leading to voltage stability considerations; see e.g.,~\cite{Bai_2022} and references therein. Future work will consider pertinent approximations and linearizations of the power flow equations for developing and analyzing such closed-loop control laws.

\appendix[Formulations of Power Import Objective]
\label{appendix:power}

Suppose for simplicity that a single feeder line denoted by $(\Sb,m)$ leaves the substation  $\Sb$. 
By invoking the multidimensional Ohm's law for the three-phase DN, the nodal current injections can be written using the bust admittance matrix as~\cite{Zbushafez}
\begin{align}
    \begin{bmatrix}
    {i} \\
    i_{\Sb,m} 
    \end{bmatrix} &= \begin{bmatrix}
                      \mathbf{Y} \\
                      \mathbf{Y}_{\Sb}
                      \end{bmatrix} {v} \label{eq:nodalinjections}
\end{align}
where, $i\in\mathbb{C}^{3N}$  collects nodal current injections $i_n$ for all buses $n \in \Nb_{+}$ (excluding slack bus) and $i_{\Sb,m}\in\mathbb{C}^3$ is the current injecton at the slack bus. Vector $v\in\mathbb{C}^{3(N+1)}$ collects the voltages for all the buses including the slack bus. The matrices $\mathbf{Y}$ and $\mathbf{Y}_{\Sb}$ are constructed using the series models of distribution lines, SVRs (grounded-wye grounded-wye), and the transformers.   

Using~\eqref{eq:nodalinjections}, the current injection at the slack bus $i_{\Sb, m}$ can be written as 
\begin{align}
\label{eq:currentinjection}
    i_{\Sb, m} &= \mathbf{Y}_{\Sb} {v} 
\end{align}

Because it is assumed that a single line leaves the slack bus, the summation is dropped from~\eqref{eq:powerimport}. 
Introducing~\eqref{eq:currentinjection} into~\eqref{eq:powerimport} yields two equivalent ways to write the power import objective function:
\begin{align}
\label{eq:zbusandpowerimportobj}
    C & = \mathrm{Re}\left\{\bm{1}^{\top}_{3} \mathrm{diag}(v_{\Sb} \bar{i}_{\Sb,m})\right\} = \mathrm{Re}\left\{\bm{1}^{\top}_{3} \mathrm{diag}(v_{\Sb} \bar{{v}}\bar{\mathbf{Y}}_{\Sb})\right\}.
\end{align}
\IEEEtriggeratref{16}
\bibliographystyle{IEEEtran}
\bibliography{IEEEabrv.bib, refconf.bib}

\begin{thebibliography}{10}
\providecommand{\url}[1]{#1}
\csname url@samestyle\endcsname
\providecommand{\newblock}{\relax}
\providecommand{\bibinfo}[2]{#2}
\providecommand{\BIBentrySTDinterwordspacing}{\spaceskip=0pt\relax}
\providecommand{\BIBentryALTinterwordstretchfactor}{4}
\providecommand{\BIBentryALTinterwordspacing}{\spaceskip=\fontdimen2\font plus
\BIBentryALTinterwordstretchfactor\fontdimen3\font minus
  \fontdimen4\font\relax}
\providecommand{\BIBforeignlanguage}[2]{{%
\expandafter\ifx\csname l@#1\endcsname\relax
\typeout{** WARNING: IEEEtran.bst: No hyphenation pattern has been}%
\typeout{** loaded for the language `#1'. Using the pattern for}%
\typeout{** the default language instead.}%
\else
\language=\csname l@#1\endcsname
\fi
#2}}
\providecommand{\BIBdecl}{\relax}
\BIBdecl

\bibitem{LA100_techreport}
J.~Cochran and D.~Paul, ``{The Los Angeles $100\%$ Renewable Energy Study},''
  National Renewable Energy Laboratory, Tech. Rep., 2021.

\bibitem{Alsaleh_2021}
I.~Alsaleh and L.~Fan, ``{Multi-Time Co-optimization of Voltage Regulators and
  Photovoltaics in Unbalanced Distribution Systems},'' \emph{IEEE Transactions
  on Sustainable Energy}, vol.~12, no.~1, pp. 482--491, 2021.

\bibitem{Shibani_2019}
S.~Ghosh, F.~Ding, J.~Simpson, T.~Harris, M.~Baggu, H.~G. Aghamolki, and
  W.~Ren, ``Techno-economic analysis for grid edge intelligence: A preliminary
  study on smart voltage regulator controls,'' in \emph{Proc. Power Energy
  Society Innovative Smart Grid Technologies Conference (ISGT)}, Washington,
  DC, USA, Feb. 2019, pp. 1--5.

\bibitem{nagarajan2018studies}
A.~Nagarajan, M.~H. Coddington, D.~Brown, S.~Hassan, L.~Franciosa, and
  E.~Sison-Lebrilla, ``{Studies on the Effects of High Renewable Penetrations
  on Driving Point Impedance and Voltage Regulator Performance: National
  Renewable Energy Laboratory/Sacramento Municipal Utility District Load Tap
  Changer Driving Point Impedance Project},'' National Renewable Energy
  Lab.(NREL), Golden, CO (United States), Tech. Rep., 2018.

\bibitem{Wu2017}
W.~Wu, Z.~Tian, and B.~Zhang, ``{An Exact Linearization Method for {OLTC} of
  Transformer in Branch Flow Model},'' \emph{{IEEE} Trans. Power Syst.},
  vol.~32, no.~3, pp. 2475--2476, 2017.

\bibitem{Savasci_2021}
A.~Savasci, A.~Inaolaji, S.~Paudyal, and S.~Kamalasadan, ``{Efficient
  Distribution Grid Optimal Power Flow with Discrete Control of Legacy Grid
  Devices},'' in \emph{Proc. Power Energy Society General Meeting (PESGM)},
  Washington, DC, USA, Jul. 2021, pp. 1--5.

\bibitem{gan2014convex}
L.~Gan and S.~H. Low, ``Convex relaxations and linear approximation for optimal
  power flow in multiphase radial networks,'' in \emph{Proc. Power Systems
  Computation Conference}, Wroclaw, Poland, 2014, pp. 1--9.

\bibitem{Jha2019}
R.~R. Jha, A.~Dubey, C.~C. Liu, and K.~P. Schneider, ``{Bi-Level Volt-VAR
  Optimization to Coordinate Smart Inverters With Voltage Control Devices},''
  \emph{{IEEE} Trans. Power Syst.}, vol.~34, no.~3, pp. 1801--1813, 2019.

\bibitem{RobbinsOPF}
B.~A. Robbins and A.~D. Domínguez-García, ``Optimal reactive power dispatch
  for voltage regulation in unbalanced distribution systems,'' \emph{{IEEE}
  Trans. Power Syst.}, vol.~31, no.~4, pp. 2903--2913, 2016.

\bibitem{Arnold_2016}
D.~B. Arnold, M.~Sankur, R.~Dobbe, K.~Brady, D.~S. Callaway, and A.~Von~Meier,
  ``{Optimal dispatch of reactive power for voltage regulation and balancing in
  unbalanced distribution systems},'' in \emph{Proc. IEEE Power and Energy
  Society General Meeting (PESGM)}, Boston, MA, USA, Jul. 2016, pp. 1--5.

\bibitem{Inaolaji_2021}
A.~Inaolaji, A.~Savasci, S.~Paudyal, and S.~Kamalasadan, ``{Accuracy of
  Phase-Decoupled and Phase-Coupled Distribution Grid Power Flow Models},'' in
  \emph{Proc. IEEE Power Energy Society Innovative Smart Grid Technologies
  Conference (ISGT)}, Washington, DC, USA, Feb. 2021, pp. 1--5.

\bibitem{Dhople_2015}
S.~V. Dhople, S.~S. Guggilam, and Y.~C. Chen, ``Linear approximations to ac
  power flow in rectangular coordinates,'' in \emph{Proc. Annual Allerton Conf.
  Communication, Control, and Computing (Allerton)}, Monticello, IL, USA, Sep.
  2015, pp. 211--217.

\bibitem{Li_2018}
C.~Li, V.~Rasouli~Disfani, Z.~Pecenak, S.~Mohajeryami, and J.~Kleissl,
  ``Optimal {OLTC} voltage control scheme to enable high solar penetrations,''
  \emph{Electric Power Systems Research}, vol. 160, pp. 318--326, 2018.

\bibitem{Li_2020}
C.~Li, V.~R. Disfani, H.~V. Haghi, and J.~Kleissl, ``Coordination of {OLTC} and
  smart inverters for optimal voltage regulation of unbalanced distribution
  networks,'' \emph{Electric Power Systems Research}, vol. 187, p. 106498,
  2020.

\bibitem{Garcia_2018}
V.~García, C.~González-Morán, and P.~Arboleya, ``Optimal tap configuration
  for step-voltage regulators applied to residential feeders,'' in \emph{Proc.
  IEEE Power Energy Society General Meeting (PESGM)}, Portland, OR, USA, Aug.
  2018, pp. 1--5.

\bibitem{Robbins2016}
B.~A. Robbins, S.~Member, H.~Zhu, and A.~D. Dom{\'{i}}nguez-garc{\'{i}}a,
  ``{Optimal Tap Setting of Voltage Regulation Transformers in Unbalanced
  Distribution Systems},'' \emph{{IEEE} Trans. Power Syst.}, vol.~31, no.~1,
  pp. 256--267, Jan. 2016.

\bibitem{Liu_2018}
Y.~Liu, J.~Li, L.~Wu, and T.~Ortmeyer, ``{Chordal Relaxation Based ACOPF for
  Unbalanced Distribution Systems With DERs and Voltage Regulation Devices},''
  \emph{{IEEE} Trans. Power Syst.}, vol.~33, no.~1, pp. 970--984, 2018.

\bibitem{Liu_2019}
Y.~Liu, J.~Li, and L.~Wu, ``Coordinated optimal network reconfiguration and
  voltage regulator/der control for unbalanced distribution systems,''
  \emph{{IEEE} Trans. Smart Grid}, vol.~10, no.~3, pp. 2912--2922, 2019.

\bibitem{Hafez_PSCC}
M.~Bazrafshan, N.~Gatsis, and H.~Zhu, ``{Optimal Tap Selection of Step-Voltage
  Regulators in Multi-Phase Distribution Networks},'' in \emph{Proc. Power
  Systems Computation Conference (PSCC)}, Dublin, Ireland, Jun. 2018, pp. 1--7.

\bibitem{Hafez_OPF_SVR}
------, ``{Optimal Power Flow With Step-Voltage Regulators in Multi-Phase
  Distribution Networks},'' \emph{{IEEE} Trans. Power Syst.}, vol.~34, no.~6,
  pp. 4228--4239, 2019.

\bibitem{Garces_2016}
A.~Garces, ``A linear three-phase load flow for power distribution systems,''
  \emph{{IEEE} Trans. Power Syst.}, vol.~31, no.~1, pp. 827--828, 2016.

\bibitem{Bernstein_2017}
A.~Bernstein and E.~Dall'Anese, ``Linear power-flow models in multiphase
  distribution networks,'' in \emph{Proc.IEEE PES Innovative Smart Grid
  Technologies Conference (ISGT-Europe)}, Turin, Italy, Sep. 2017, pp. 1--6.

\bibitem{Huang_2021}
J.~Huang, B.~Cui, X.~Zhou, and A.~Bernstein, ``A generalized lindistflow model
  for power flow analysis,'' in \emph{Proc. IEEE Conf. Decision and Control
  (CDC)}, Austin, TX, USA, Dec. 2021, pp. 3493--3500.

\bibitem{Girigoudar_2021}
K.~Girigoudar and L.~A. Roald, ``Linearized three-phase optimal power flow
  models for distribution grids with voltage unbalance,'' in \emph{Proc. IEEE
  Conference on Decision and Control (CDC)}, Austin, TX, USA, Dec. 2021, pp.
  4214--4221.

\bibitem{Kersting_2018}
W.~H. Kersting, \emph{Distribution System Modeling and Analysis}, 4th~ed.\hskip
  1em plus 0.5em minus 0.4em\relax Boca Raton, FL, USA: CRC Press, 2018.

\bibitem{Zbushafez}
M.~Bazrafshan and N.~Gatsis, ``{Comprehensive Modeling of Three-Phase
  Distribution Systems via the Bus Admittance Matrix},'' \emph{{IEEE} Trans.
  Power Syst.}, vol.~33, no.~2, pp. 2015--2029, Mar. 2018.

\bibitem{sankur2016linearized}
M.~D. Sankur, R.~Dobbe, E.~Stewart, D.~S. Callaway, and D.~B. Arnold, ``A
  linearized power flow model for optimization in unbalanced distribution
  systems,'' \emph{arXiv preprint arXiv:1606.04492}, 2016.

\bibitem{Lofberg2004}
J.~L{\"{o}}fberg, ``Yalmip : A toolbox for modeling and optimization in
  matlab,'' in \emph{Proc. of the CACSD Conference}, Taipei, Taiwan, 2004.

\bibitem{mosek}
\BIBentryALTinterwordspacing
M.~ApS, \emph{The MOSEK optimization toolbox for MATLAB manual. Version 9.3.},
  2021. [Online]. Available: \url{http://docs.mosek.com/9.3/toolbox/index.html}
\BIBentrySTDinterwordspacing

\bibitem{Grant2014}
M.~Grant and S.~Boyd, ``{CVX}: Matlab software for disciplined convex
  programming, version 2.1,'' \url{http://cvxr.com/cvx}, Mar. 2014.

\bibitem{ConvZBus}
M.~Bazrafshan and N.~Gatsis, ``Convergence of the {Z}-bus method for
  three-phase distribution load-flow with {ZIP} loads,'' \emph{{IEEE} Trans.
  Power Syst.}, vol.~33, no.~1, pp. 153--165, Jan. 2018.

\bibitem{Stuhlmacher_2020}
A.~{Stuhlmacher} and J.~L. {Mathieu}, ``Chance-constrained water pumping to
  manage water and power demand uncertainty in distribution networks,''
  \emph{Proc. IEEE}, vol. 108, no.~9, pp. 1640--1655, 2020.

\bibitem{Bai_2022}
B.~Cui, A.~S. Zamzam, G.~Cavraro, and A.~Bernstein, ``Efficient region of
  attraction characterization for control and stabilization of load tap changer
  dynamics,'' \emph{{IEEE} Trans. Control Netw. Syst.}, pp. 1--1, 2022, early
  access.

\end{thebibliography}
\end{document}